\documentclass[%
reprint,
showpacs,
 amsmath,amssymb,
 aps,
 prl,
]{revtex4-1}

\usepackage{graphicx}
\usepackage{dcolumn}
\usepackage{bm}

\begin{document}

\title{Evidence for universal conductance correction in a tunable strongly coupled nanogranular metal}

\author{Roland Sachser}
 \email{sachser@physik.uni-frankfurt.de}
\author{Fabrizio Porrati}%
\author{Christian H. Schwalb}%
\author{Michael Huth}%
\affiliation{%
 Physikalisches Institut, Goethe-Universit\"at, Max-von-Laue-Str. 1, 60438 Frankfurt am Main, Germany}%

\date{\today}

\begin{abstract}
We present temperature-dependent conductivity data obtained on a sample set of nanogranular Pt-C with finely tuned inter-grain tunnel coupling strength $g$. For samples in the strong-coupling regime $g>g_C$, characterized by a finite conductivity for $T\rightarrow 0$, we find a logarithmic behavior at elevated temperatures and a crossover to a $\sqrt{T}$-behavior at low temperatures over a wide range of coupling strengths $g_C\approx 0.25 < g \leq 3$. The experimental observation for $g>1$ is in very good agreement with recent theoretical findings on ordered granular metals in three spatial dimensions. The results indicate a validity of the predicted universal conductivity behavior that goes beyond the immediate range of the approach used in the theoretical derivation.
\end{abstract}

\pacs{73.23.Hk, 71.30.+h, 73.63.Bd, 81.07.Bc}

\maketitle


Nanogranular metals represent model systems for the study of the interplay of electronic correlations, quantum confinement effects and disorder.
They are formed by nanometer-sized metallic grains which are embedded in an insulating, polarizable matrix. The intergrain coupling strength $g$, normalized to the quantum conductance $2e^2/\hslash$ for one spin direction, determines their electronic properties.
For $g > g_C$ metallic behavior ensues which is characterized by a non-zero electrical conductivity for $T \rightarrow 0$, whereas for $g < g_C$ the granular metal becomes insulating at low temperatures. The critical coupling strength $g_C=(1/2\pi d)\ln{(E_C/\delta)}$ is logarithmically dependent on the ratio of the averaged Coulomb charging energy $E_C$ of the grains and the averaged level spacing $\delta$ within a grain at the chemical potential, see e.\ g.\  \cite{Beloborodov2007}. $d$ denotes the spatial dimension.
Early studies on this material class date back to the 1970ies and 1980ies, as has been reviewed theoretically and experimentally \cite{Abeles1975, Sheng1992}.
Recently, a renewed interest in these materials can be observed for several reasons.
It has become clear that high-temperature superconductors are intrinsically disordered which is assumed to lead to a self-induced granularity of the superfluid density \cite{Imry2008, Dubi2007} for which also experimental evidence has been found, e.g. in low-temperature scanning tunneling spectroscopy \cite{Pan2001}.
It is speculated whether this may be relevant with regard to the observation of the pseudogap phenomenon \cite{Timusk1999, Alvarez2005}.
A similar observation has been made for thin granular metal films which exhibit a thickness-dependent superconductor-to-insulator transition associated with the evolution of a granular electronic density distribution \cite{Frydman2002}.
Interestingly, and also as an impetus for the work presented here, new theoretical findings for nanogranular metals in the strong inter-grain coupling limit predict distinctively different transport behaviors in a high-energy regime ($k_BT>g\delta$), dominated by single-grain physics, and a low-energy regime ($k_BT<g\delta$) which shows strong similarities to a homogeneously disordered metal \cite{Beloborodov2003, Beloborodov2007}.
The low-energy regime has been associated with the notion of a granular Fermi liquid \cite{Beloborodov2004}.

We report measurements of the temperature-dependent electrical conductivity of a series of identically prepared nanogranular metal samples whose inter-grain coupling strength has been subject to a continuous tuning by electron irradiation.
We have been able to very finely tune the coupling strength to the metallic side of the metal-insulator transition and find clear experimental proof for the separation into a high and low energy regime, as was theoretically predicted \cite{Beloborodov2003}. We introduce a robust procedure for obtaining the inter-grain coupling strength $g$ from normalized temperature-dependent conductivity data and find the same universal temperature dependence of the conductivity going beyond the immediate validity range of the theory $(g \gg 1)$ into the range $g_C \approx 0.25 < g \leq 3$.


The samples were prepared by focused electron beam induced deposition (FEBID), see e.~g.\ \cite{Utke2008}, followed by an additional electron beam irradiation treatment.
We employed a FEI Nova Nanolab 600 SEM with a Schottky-type emitter. The metal-organic platinum precursor $(CH_3)_3CH_3C_5H_4Pt$ was used and supplied close the electron beam focal point on the substrate surface by means of a capillary of $0.5~mm$ diameter. $120~nm$ Au/Cr electrodes were defined on the Si(100)/SiO$_2$ (300~nm) substrate by UV lithography on the substrate before use. During the FEBID process the electron beam is rastered over the surface, dissociating the adsorbed precursor molecules and forming the nanogranular metal deposits consisting of Pt nanocrystallites with a diameter of $3.2~nm\pm 0.8~nm$ \cite{Teresa2009} embedded in a carbon matrix. An acceleration voltage of $5~kV$ and a beam current of $1.6~nA$ were used. Under these beam conditions the metal content amounts to $22~at\%$ for the as-grown samples.

An in-situ measurement setup, which allows us to measure the conductance of the samples during the deposition process, was employed which allowed us to verify a very high reproducibility and comparability of the electrical properties in sample preparation \cite{Porrati2009}. Several samples with a thickness of $82~nm$ each were prepared by means of FEBID on one substrate under identical conditions.

Following the deposition process and a waiting time to ensure that no more precursor molecules were in the vacuum chamber, an electron beam irradiation treatment without precursor gas flux was applied to the samples. The same beam parameters as during the deposition process were used, but the irradiation time and, respectively, the irradiation dose was varied. This allowed us to finely tune the samples' conductance to the desired values with a very large degree of controllability. Up to the corresponding irradiation time, i.\ e.\  irradiation dose, the measurements show similar behavior for the different samples. Further details about this approach can be found in \cite{Schwalb2010, Porrati2011}.

Temperature-dependent transport measurements were performed in the range of $1.5~K$ to $260~K$ using a $^4$He-cryostat equipped with a dynamic variable temperature insert.
For conductivity measurements Keithley SourceMeters, Model No. 2400 and 2636A, were used to apply a fixed bias voltage of $10~mV$ to the samples resulting in an electric field of no more than $25~V/cm$. Under these bias voltage conditions the measurements on the metallic samples were taken in the linear regime as was checked by the current-voltage characteristics measured at the lowest temperature. Self-heating effects of the samples could thus be excluded. The data were taken in two-probe geometry. Contact and wiring resistances could be neglected as was verified by independent three and four-probe measurements.

The time between venting the SEM and mounting the samples to the cryostat was kept as short as possible. Further measurements, such as atomic force microscopy measurements (AFM) in non-contact mode to determine the height of the samples, and EDX measurements to analyze their composition, were performed after the temperature-dependent conductivity measurements. Thereby aging effects were kept as small as possible and the uncontrolled influence of additional irradiation caused by EDX before taking the transport data was avoided.


In Fig.~1(a) we present the normalized temperature-dependent conductivity of samples irradiated with a dose up to $6.72~\mu C/\mu m^{2}$. Apparently, the transport behavior is strongly altered by varying the irradiation dose. As-grown samples show insulating behavior indicated by the temperature-dependent conductivity following a stretched exponential behavior, corresponding to correlated variable range hopping \cite{Porrati2011,Efros1975}. This is a well-known behavior for granular metals prepared by FEBID for different precursors \cite{Huth2009,Tsukatani2005,Teresa2009}.
This insulating characteristics is observed for all samples exposed to doses below about $0.32~\mu C/\mu m^{2}$. With further increase of the irradiation dose the conductivity also increases and the samples pass a insulator-to-metal transition.
For the sample exposed to the highest irradiation dose of $6.72~\mu C/\mu m^{2}$ the resistivity even increases with increasing temperature, as would be expected for a conventional metal. For the set of samples irradiated from $0$ to $6.72~\mu C/\mu m^{2}$ the conductivity at room temperature varies over three orders of magnitude ranging in absolute values from $16~\Omega^{-1}m^{-1}$ to $1.25 \cdot 10^{5}~\Omega^{-1}m^{-1}$. 

In order to establish the transition between metallic and insulating behavior reliably we analyzed the logarithmic derivative $w=dln\sigma /dlnT$ of the conductivity which defines a more accurate criterion to distinguish between metallic and insulating behavior than simply extrapolating $\sigma (T)$ to $T=0$ \cite{Mobius1989}.
Metallic behavior leads to a vanishing $w$ for approaching $T=0$, insulating behavior instead is indicated by a constant or divergent value for $w$.
In Fig.~1(b) the logarithmic derivative is shown for three samples irradiated with doses in the range $0.32~\mu C/\mu m^{2}$ to $0.64~\mu C/\mu m^{2}$.
$w$ diverges for the sample with $0.32~\mu C/\mu m^{2}$, indicating its still insulating behavior. For samples exposed to a larger dose $w$ tends to 0 for $T \rightarrow 0$, so they are metallic.
We will now focus on the metallic samples exposed to doses $\geq 0.48~\mu C/\mu m^{2}$.

\begin{figure}
\includegraphics[angle=-90,width=\columnwidth]{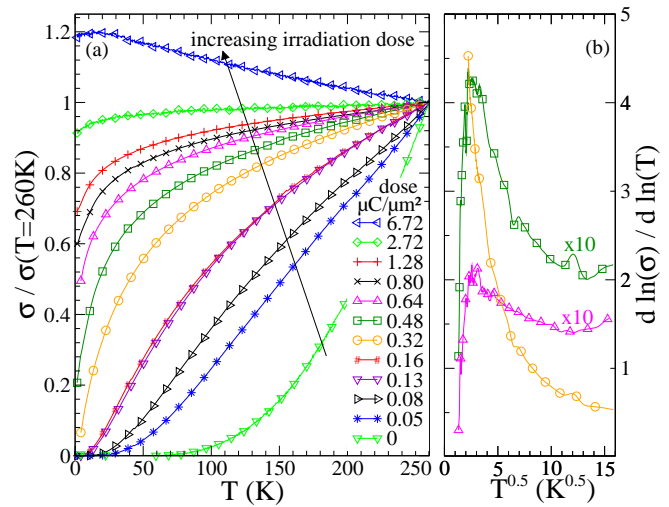}%
\caption{\label{Normalized temperature-dependent conductivity and logarithmic derivative} (Color online) (a) Normalized temperature-dependent conductivity for samples irradiated with a dose up to $6.72~\mu C/\mu m^{2}$. (b) Logarithmic derivative $w=dln\sigma /dlnT$ for samples near the insulator-to-metal transition irradiated witch a dose from $0.32~\mu C/\mu m^{2}$ to $0.64~\mu C/\mu m^{2}$. The sample irradiated exposed to $0.32~\mu C/\mu m^{2}$ shows insulating behavior, the other two samples tend to be metallic.}
\end{figure}

In Fig.~2(a) the logarithmic temperature dependence of the normalized conductivity is plotted for metallic samples irradiated with a dose from  $0.48~\mu C/\mu m^{2}$ to $2.72~\mu C/\mu m^{2}$. This transport behavior is observable starting from about $6~K$ to $18~K$, depending on the sample, up to $260~K$, the highest temperature measured with our setup. In Fig.~2(b) we show the same data vs.\ $\sqrt{T}$ and find linear behavior from the lowest temperature measured ($1.5~K$) up to a temperature of $10~K$ to $25~K$ depending on the sample. We exclude the sample irradiated with $6.72~\mu C/\mu m^{2}$ from the following analysis because of its positive temperature coefficient of resistance which clearly indicates that the percolation threshold has already been passed for this sample.

\begin{figure}
\includegraphics[angle=-90,width=\columnwidth]{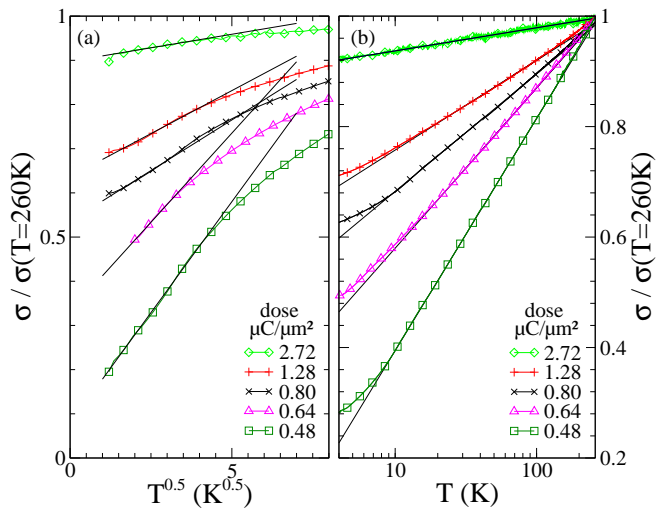}%
\caption{\label{log and sqrt behavior} (Color online) (a) Logarithmic and (b) square-root temperature behavior for samples irradiated with a dose in the range from  $0.48~\mu C/\mu m^{2}$ to $2.72~\mu C/\mu m^{2}$. The data sets are represented by the solid lines with the equally spaced symbols provided for ease of discrimination.}
\end{figure}

We now turn to discussing possible reasons for these observations. Beloborodov et al. proposed a theory of ordered granular metals in the metallic, i.\ e.\ the strong inter-grain coupling regime \cite{Beloborodov2003}. Two distinct transport regimes are predicted which lead to corrections to the diffusive intra-grain conductivity $\sigma_0$ in the following way:
\begin{equation}
\sigma = \sigma _{0} + \delta \sigma _{1} + \delta \sigma _{2} \quad .
\label{eq1}
\end{equation}
$\delta\sigma_1$ represents the correction in a high energy regime ($k_BT>g\delta$) which is dominated by the granular structure and incoherent tunneling processes. $\delta\sigma_2$ stands for the correction in a low energy regime ($k_BT<g\delta$) in which coherent electron motion is established. The low energy regime shows similarities to homogeneously disordered metals and has been associated with the notion of a granular Fermi liquid \cite{Beloborodov2004}. The temperature-dependent corrections in leading order are as follows \cite{Beloborodov2003}
\begin{equation}
\frac{\delta \sigma _{1}}{\sigma _{0}} = - \frac{1}{6 \pi g} ln \left[ \frac{gE_{C}}{max(k_{B}T,g\delta)} \right]
\label{eq2}
\end{equation}
and
\begin{equation}
\frac{\delta \sigma _{2}}{\sigma _{0}} = \frac{1.83}{12 \pi ^2 g} \sqrt{\frac{k_{B}T}{g\delta}}\quad .
\label{eq3}
\end{equation}
assuming $d=3$ in the present case.

First experimental indications of a logarithmic temperature correction of the conductivity was reported by Rotkina et al.\ for one platinum-containing sample prepared by FEBID which had been subjected to an additional heat treatment \cite{Rotkina2005}. This behavior was tentatively attributed to the high energy regime and the first correction term. Our samples show both, a logarithmic temperature dependence and also a crossover behavior to a distinctly different low-temperature behavior which is following a $\sqrt{T}$-dependence within the limited temperature range $T>1.5~K$ available in our setup.

\begin{figure}
\includegraphics[angle=-90,width=\columnwidth]{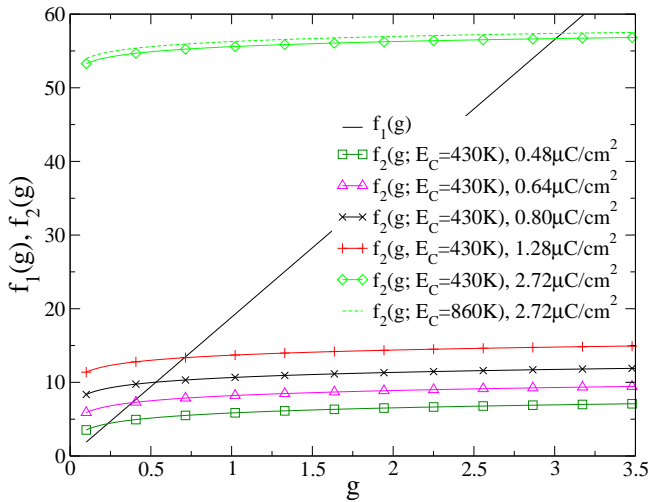}%
\caption{\label{fig3} (Color online) Plot of left and right part of Eq.~\ref{eq5} denoted as $f_1(g)$ and $f_2(g)$, respectively. The intersection point provides the inter-grain coupling strength for the respective nanogranular metal. The insensitivity of the derived values for $g$ is exemplarily shown for the sample exposed to the largest dose assuming two very different values for the Coulomb charging energy as indicated.}
\end{figure}

For further analysis it is mandatory to quantify the inter-grain coupling strength $g$ for our sample set for the following reason. The theoretically predicted behavior in the temperature regime $k_BT > g\delta$ stems from the renormalization of the tunnel coupling $g$ caused by Coulomb correlations. Perturbation theory in $1/g$ generalized by a renormalization group approach has been used to cover the range $g\gg 1$ \cite{Beloborodov2003}. Here we use the data that follow a $\ln{T}$ behavior to derive $g$ in the following way. According to Eq.~\ref{eq2} for $d=3$ the normalized conductivity $\sigma(T)/\sigma(T_n)$ ($T_n=260~K$ here) follows a linear behavior
\begin{equation}
\frac{\sigma(T)}{\sigma(T_n)} = a + m\ln{T[K]}
\label{eq4}
\end{equation}
After some simple algebra and using our experimentally determined values for $a$ and $m$ taken from Fig.~\ref{log and sqrt behavior}(a) the following condition can be derived
\begin{equation}
2\pi dg = a/m + \ln{\left[(gE_C/k_B) [K]\right]}
\label{eq5}
\end{equation}
This condition equation depends only weakly on the magnitude of the Coulomb charging energy which we can estimate from the grain size of our samples. The averaged grain size $D$ amounts to $3.2~nm$, as determined by transmission electron microscopy for as-grown samples \cite{Teresa2009}, and we take into account an approximate size increase of $20~\%$ caused by the electron irradiation treatment \cite{Porrati2011}. This leads to an average grain diameter of $3.9~nm$ which results in a charging energy of $E_C/k_B = e^2/2k_BC = e^2/8k_B\pi\epsilon_0\epsilon_rD \approx 430~K$ with a relative permittivity of $\epsilon_r = 5$. Since neither $D$ nor $\epsilon_r$ are exactly known we have to allow for a range of possible $E_C$ values. This uncertainty does only cause very weak changes in the deduced coupling strength as can be seen in Fig.~\ref{fig3} which shows the left part, denoted as $f_1(g)$, and right part, denoted as $f_2(g)$, of Eq.~\ref{eq5} for the samples in the strong inter-grain coupling regime. The fit parameters $a$ and $m$ referring to the logarithmic temperature dependence shown in Fig.~\ref{log and sqrt behavior}(a) have been used. The intersection points give the inter-grain tunnel coupling strength $g$ for each sample.

Apparently, only the sample exposed to the largest dose of $2.72~\mu C/cm^2$ having $g\approx 3$ is in the strong-coupling regime within the immediate validity range of the theory \cite{Beloborodov2003}. The estimated average level spacing $\delta = 1/(V \cdot N_f)$ amounts to $2.8~K$, with V as the volume of an individual grain of diameter $3.9~nm$ and
$N_f \approx 2/(eV \cdot atom)$ \cite{Muller1971} the density of states at the Fermi level. Using this and the coupling strength $g=3$ we obtain $T^* = 8.4~K$ as crossover temperature between the high- and low-temperature regime. This is in good correspondence with the observed crossover temperature taken from the respective fits in Fig.~\ref{log and sqrt behavior}(a) and (b) which amounts to about $10~K$.

For the other samples it can be stated that these also fall onto the metallic side of the insulator-to-metal transition since $g_C \approx 0.25$. Apparently, the logarithmic temperature dependence and crossover behavior at low temperatures is a robust and universal feature over a large range of coupling strength reaching well below the expected validity range of the theory. It remains to be seen whether the aspect of disorder, mainly caused by the inter-grain distance distribution, does help to stabilize the metallic regime against the correlation effects which favor localization of the electrons.


In conclusion, we have presented experimental evidence for a universal low-temperature behavior of the electrical conductivity of a three-dimensional nanogranular metal in the strong inter-grain coupling regime. This has become possible by introducing a new approach for tuning the inter-grain coupling strength employing an electron irradiation treatment under continuous monitoring of the sample conductance. We find the theoretically predicted logarithmic temperature dependence of the conductivity at high temperatures followed by a crossover behavior to coherent electron motion at low temperature, as indicated by a $\sqrt{T}$-behavior. Our results suggest that the theoretically predicted behavior, derived in the strong-coupling limit $g\gg 1$, has in fact a larger validity range than could be expected. Disorder in the inter-grain coupling strength may be a reason for this observation. As a consequence of the large degree of control immanent to the FEBID process employed in sample preparation, extension of these studies to two-dimensional, as well as one-dimensional nanogranular metals appear feasible \cite{Sachser2009, Porrati2010}. Moreover, the continuous tunability of the inter-grain coupling strength will allow future investigations in the immediate neighborhood of the insulator-to-metal transition. It will also allow for establishing a phase diagram of the charge carrier dynamics \cite{Beloborodov2004} for the Pt-C system.


Financial support by the Beilstein-Institut, Frankfurt/Main, Germany, within the research collaboration NanoBiC and
by the NanoNetzwerkHessen (NNH) is gratefully acknowledged.



\begin{thebibliography}{}
\bibitem{Beloborodov2007} I. S. Beloborodov, A. V. Lopatin, V. M. Vinokur, and K. B. Efetov, Rev. Mod. Phys. \textbf{79}, 469 (2007).
\bibitem{Abeles1975} B. Abeles, P. Sheng, M. D. Coutts, and Y. Arie, Adv. Phys. \textbf{24}, 407 (1975).
\bibitem{Sheng1992} P. Sheng, Philos. Mag. B \textbf{65}, 357 (1992).
\bibitem{Imry2008} Y. Imry, M. Strongin, and C. C. Homes, Physica C \textbf{468}, 288 (2008).
\bibitem{Dubi2007} Y. Dubi, Y. Meir and Y. Avishai, Nature \textbf{449}, 876 (2007).
\bibitem{Pan2001} S. Pan et al., Nature \textbf{413}, 282 (2001).
\bibitem{Timusk1999} T. Timusk and B. Statt, Rep. Progr. Physi. \textbf{62}, 61 (1999).
\bibitem{Alvarez2005} G. Alvarez, M. Mayr, A. Moreo, and E. Dagotto, Phys. Rev. B \textbf{71}, 014514 (2005).
\bibitem{Frydman2002} A. Frydman, O. Naaman, and R. C. Dynes, Phys. Rev. B \textbf{66}, 052509 (2002).
\bibitem{Beloborodov2003} I. S. Beloborodov, K. B. Efetov, A. V. Lopatin, and V. M. Vinokur, Phys. Rev. Lett. \textbf{91}, 246801 (2003).
\bibitem{Beloborodov2004} I. S. Beloborodov, A. V. Lopatin, and V. M. Vinokur, Phys. Rev. B \textbf{70}, 205120 (2004).
\bibitem{Utke2008}  I. Utke, P. Hoffmann, and J. Melngailis, J. Vac. Sci. Technol. B \textbf{11}, 2386 (2008).
\bibitem{Teresa2009} J. M. De Teresa, R. C\'ordoba, A. Fern\'andez-Pacheco, O. Montero, P. Strichovanec, M. R. Ibarra, J. Nanomat. (2009), Article Number 936863.
\bibitem{Porrati2009} F. Porrati, R. Sachser, and M. Huth, Nanotechnology \textbf{20}, 195301 (2009).
\bibitem{Schwalb2010} Ch. Schwalb, C. Grimm, M. Baranowski, R. Sachser, F. Porrati, H. Reith, P. Das, J. Mu\"uller, F. V\"olklein, A. Kaya, and M. Huth, Sensor \textbf{10}, 9847 (2010).
\bibitem{Porrati2011} F. Porrati, R. Sachser, C. H. Schwalb, A. S. Frangakis, and M. Huth, J. Appl. Phys. \textbf{109}, 063715 (2011).
\bibitem{Efros1975} A. L. Efros and B. I. Shklovskii, J. Phys. C \textbf{8}, L49 (1975).
\bibitem{Huth2009} M. Huth, D. Klingenberger, Ch. Grimm, F. Porrati, and R. Sachser, New J. Phys. \textbf{11}, 033032 (2009).
\bibitem{Tsukatani2005} Y. Tsukatani, N. Yamasaki, K. Murakami, F. Wakaya, and M. Takai, Jpn. J. Appl. Phys., Part 1 \textbf{44}, 5683 (2005).
\bibitem{Mobius1989} A. M\"obius, Phys. Rev. B \textbf{40}, 4194 (1989).
\bibitem{Rotkina2005} L. Rotkina, S. Oh, J. N. Eckstein, and S. V. Rotkin, Phys. Rev. B. \textbf{72}, 233407 (2005)
\bibitem{Muller1971} F. M. M\"uller, J. W. Garland, M. H. Cohen, and K. H. Bennemann, Annals of Physics \textbf{67}, 19-57 (1971)
\bibitem{Sachser2009} R. Sachser, F. Porrati, and M. Huth, Phys. Rev. B \textbf{80}, 195416 (2009).
\bibitem{Porrati2010} F. Porrati , R. Sachser , M. Strauss , I. Andrusenko , T. Gorelik , U. Kolb , L. Bayarjargal , B. Winkler, and M. Huth, Nanotechnology \textbf{21}, 375302 (2010).
\end{thebibliography}

\end{document}